# Fractional Order Shapiro Steps in Superconducting Nanowires

R.C. Dinsmore III, Myung-Ho Bae, and A. Bezryadin

*Department of Physics*

*University of Illinois at Urbana-Champaign,*

*Urbana, Illinois 61801-3080, USA*

*We expose superconducting nanowires to microwave radiation in order to study phase lock-in effects in quasi one dimensional superconductors. For sufficiently high microwave powers a resistive branch with Shapiro steps appears in the voltage-current characteristics. At frequencies in the range of 0.9-4 GHz these steps are of integer order only. At higher frequencies steps of 1/2, 1/3, 1/4 and even 1/6 order appear. We numerically model this behavior using a multi-valued current-phase relationship (CPR) for nanowires.*

Superconducting nanowires [1,2,3] show great potential for use in novel devices because they allow one to exploit superconducting properties unique to one dimensional systems. Nanowires are also free of the decoherence problems inherent to Josephson junctions (Jj) that are related to charge fluctuations in the dielectric barrier of the tunnel junction [4]. Examples of interesting and potentially useful phenomenon predicted and/or occurring in thin superconducting nanowires include macroscopic quantum tunneling (MQT), [5,6,7,8] a dissipation-controlled quantum phase transition [9, 10, 11, 12, 13, 14], also known as Schmid-Bulgadaev transition [15], a quantum Kosterlitz-Thouless transition [16], a Joule-heating-driven hysteresis in current-voltage characteristics [17, 18, 19, 20], and a multi-valued character of the current-phase relationship (CPR) [1, 21]. Nanowires have also been used as photon counters [22] which is an important task in radioastronomy.

Our results on Shapiro steps (Ss) in superconducting nanowires are similar to the first observations of Ss in thin film constrictions by Anderson and Dayem in that the lower integer steps are not present until higher power microwave radiation (MW) is applied[23]. Anderson and Dayem have also found half-integer resonance steps in their bridges [23]. However nanowires are a different type of system since they are quasi-one dimensional, while Dayem bridges are quasi-two dimensional. In this letter we study phase lock-in effects in quasi-one-dimensional superconducting wires and discuss the results in terms of multi-valued CPR. We find that by applying microwaves to thin wires we are able to initiate and stabilize a resistive dynamic superconducting state, i.e. a phase slip center (PSC). Shapiro steps are observed in this state and studied for MW frequencies between 0.9 and 15 GHz. At low frequencies we observe only integer order Ss, but, as the frequency is increased, fractional order steps are found, starting first with one-half steps and increasing in order to one-third, one-quarter and even one-sixth as the frequency is increased. We attribute this behavior to a multi-valued, non-sinusoidal CPR of superconducting wires and perform numerical simulations to support our conclusions.

The nanowires used in this study were fabricated using molecular templating technique [24]. We deposit droplets of a solution containing fluorinated carbon nanotubes onto a $SiN/SiO_2/Si$ substrate that has a 100 nm trench etched into the top SiN layer (with an undercut created in the underlying $SiO_2$ using HF). We then decorate the sample with a thin layer of amorphous $Mo_{76}Ge_{24}$ [25]. Some of the nanotubes lie across the trench and form nanowires connected seamlessly to the leads. These wires are examined in an scanning electron microscope (SEM) and chosen based on their apparent homogeneity and the presence of "bright spots" at the ends of the wire, which indicate that the wire is straight and properly suspended above the trench (see inset Fig. 1a) [26]. These samples are current-biased in a 4-probe configuration and their transport properties are measured in a $^3$He cryostat. The cryostat is equipped with silver paste and copper powder filters, held at temperature $T$=300 mK and RF $\pi$-filters at room temperature to filter out electromagnetic noise in the DC lines. The MW is fed in through a stainless steel coaxial microwave line with a -10 dB attenuator kept at 4K and a -3 dB attenuator kept at 1K, serving to limit unwanted thermal radiation and other noise. The signal from the coaxial cable is weakly coupled to the sample via a coil antenna positioned at the bottom of the sample Faraday cage.

A typical current-voltage characteristic, *V(I),* of a nanowire exposed to MW is shown in Fig. 1(a) for sample A. At zero applied MW power the *V(I)* curves are hysteretic at low temperatures (curve 1). For wires in this regime there is no measurable voltage until the current is increased beyond the critical switching current, $I_{SW}$. At this current the wire transits to the normal state which is sustained due to Joule heating by the bias current to temperatures above $T_C$ [17, 18, 19, 20]. The wire remains in this Joule-heated normal state (JNS) until the bias current is lowered below the return current, $I_R$. At this current Joule heating is no longer sufficient to heat the wire above $T_C$ and the wire "drops" back to the superconducting state. The wire resistance in the JNS, as determined from the slope of the *V(I)* curve, is very close to the normal state resistance of the wire as observed in resistance vs. temperature measurements (not shown). Also there are no Ss in the JNS indicating that there is no periodic phase evolution to synchronize with applied MW. These two results confirm that superconductivity is completely suppressed in the JNS and that it encompasses the entire length of the wire.

When the wire is subjected to microwave radiation of increasing power, the switching current is first suppressed (curve 2 in Fig.1(a)) and then an additional resistive branch, namely a PSC, appears at currents below $I_{SW}$. Curve 3 of Fig.1(a) shows the PSC branch, although it is not well pronounced since the corresponding voltage is small at this scale. If one "zooms-in" on the PSC branch one finds the curves such as those shown in Fig.1(b). These normalized *V(I)* curves show Ss indicating that there is a dynamic superconducting state in which the periodically evolving phase difference along the wire synchronizes with the external MW signal. These observations confirm that we have two qualitatively different resistive states in our superconducting nanowires: the JNS which occurs at high current and no microwave radiation and the PSC which occurs at intermediate bias current and non-zero MW radiation. Each of these states is different from the fully superconducting state (or static superconducting state) occurring at low-enough bias currents and characterized by zero voltage.

If a lock-in between the external microwave signal and the revolving of the phase of the superconducting order parameter occurs, a non-zero supercurrent through the wire becomes possible [27]. This leads to a slower increase of the voltage with the bias current. Therefore the lock-in effect occurs as a minimum on the *dV/dI*(*I*) curve. Figure 2(a) shows such resonances very clearly for sample B with 9.5 GHz MW applied. Such lock-in resonances correspond to Shapiro steps in the *V(I)* curves. In fact the central result of this Letter is summarized in Fig.2, which shows not only integer-order resonances but fractional resonances, which become stronger as the MW frequency is increased.

The PSC branch shows Ss that behave quite differently at higher frequencies than those in a Josephson junction [1]. As is apparent from Fig.1(b), one difference in nanowires is that when the MW power is increased the lower integer steps get suppressed and remain small whereas the higher order steps that appear for higher powers become more pronounced [28]. This produces what looks like a kink in the *V(I)* characteristic (see the top curve in Fig.1(b)). The second difference is that when the frequency is increased, fractional steps appear with increasing order, something that does not occur for sinusoidal CPRs. We explain these differences by assuming that CPR is strongly non-sinusoidal and multi-valued and therefore corresponds to a fast variation of the supercurrent ("jumps") at the moments of phase slips [1, 21]. The system is modeled numerically using the McCumber-Stewart resistively shunted junction model (RSJ) [29] for overdamped junctions and is described by the following phase evolution equation: $d\phi/d\tau + I_S(\phi) = i_{dc} + i_{ac}\sin(\Omega\tau) + \eta(T,\tau)$. Here $\phi$ is the phase difference across the length of the wire, $\tau$ [$=2\pi f_c t$] is the dimensionless time, $\Omega = f/f_c$, and $i_{dc}$ and $i_{ac}$ are, respectively, the DC and AC bias currents, both normalized by the critical depairing current, $I_C$. The characteristic frequency, given by $f_C = 2eI_C R/h$, accounts for the resistance of the normal current channel, $R$, and the critical current of the wire. The η(T,τ) term represents random thermal fluctuations [30, 31, 30]. We use the following CPR for the wire [1]: $I_s(\phi) = 3\sqrt{3}I_C/2[\phi/(L/\xi) - (\phi/(L/\xi))^3]$, where $L$ is the length of the wire and ξ is the coherence length. This function is valid in the limit $L >> \xi$ and in the absence of very frequent quantum phase slips [3] (see Fig. 3). Unlike Josephson Junctions, a very long wire can have a phase difference, from one end to the other, larger than π, corresponding to a state that carries a supercurrent and is physically different from states with $0 < \phi < \pi$. When a phase slip occurs in a sufficiently long nanowire, the supercurrent should decrease locally to zero (or to a negative value, in the cases when the phase difference before the phase slip was less than 2π), but the segments of the wire located far from the center of the phase slip still carry a positive supercurrent. Thus, after the phase slip is ended, the wire, if it is long enough and the initial current was high enough, still carries some nonzero positive supercurrent.

To perform numerical simulations we assume that the system can not follow the unstable branch of the CPR that begins at $I_C$, but rather undergoes a phase slip at $I_C$. The $I_C$ is defined as the maximum current on the curve representing the CPR. This process is illustrated in Fig. 3. In the model used here it is assumed that each phase slip event leads to a phase change by $2\pi$ (as is shown by red arrows in Fig.3a) and the corresponding change of the supercurrent in the wire. Numerically this amounts to allowing the phase to evolve according to the RSJ model until it reaches $\phi_{Max}(\phi_{Min})$, the value at which the supercurrent hits the critical current and then changing $\phi$ to $\phi_{Max} -(+) 2\pi$, correspondingly. Simulations were carried out using the wire length $L$ measured in the SEM. The effective coherence length $\xi$ was adjusted to better reproduce our experimental data. We find that this approach gives rather good qualitative agreement with experiment for low normalized frequencies $\Omega = 0.1–0.3$. The simulated curves are very sensitive to the parameters $\Omega$ and $\xi$ but show similar features when the typical values are used. The numerical simulations show similarly shaped $V(I)$ curves with increasing power (see Fig. 1). Our model cannot address the switching into the JNS at high bias currents or the hysteresis and missing steps at low microwave power, both features indicating the presence of meta-stable states. Fig.2 shows data obtained from sample B at various frequencies compared to numerical simulations using $L = 130$ nm and $\xi = 6$ nm. For higher frequencies the numerical simulations (Fig.2(c)) do show increasing fractional order steps, however the ½ steps do not disappear when the ⅓ steps appear, as we observe in experiments. This fact remains not understood. Numerical simulations were carried out with other CPRs and the multi-valued CPR appears to be in better agreement with the data. We therefore expect that it is the phase slip process accompanied by an abrupt change in the supercurrent in the wire that is the key to the appearance of the fractional steps.

In summary, we find that a phase slip center can be initiated and sustained by external microwave radiation. This microwave-stabilized phase slip center regime might become a new tool in photon detection. We observe increasing fractional Shapiro steps with increasing microwave frequency in superconducting nanowires. We attribute this effect to the presence of time-periodic phase slips and the non-sinusoidal and multi-valued nature of the current-phase relationship of nanowires. We apply the McCumber – Stewart model for numerical simulations and obtain results similar to the experimental ones.

Acknowledgements: The authors thank R. Giannetta for useful discussions. This work was supported by the U.S. Department of Energy under Grant No. DE-FG02–07ER46453. This work was carried out in part in the Frederick Seitz Materials Research Laboratory Central Facilities, University of Illinois, which are partially supported by the U.S. Department of Energy under Grant Nos. DE-FG02–07ER46453 and DE-FG02–07ER46471. M.H.B wants to thank the Korea Research Foundation Grants No. KRF-2006-352-C00020.

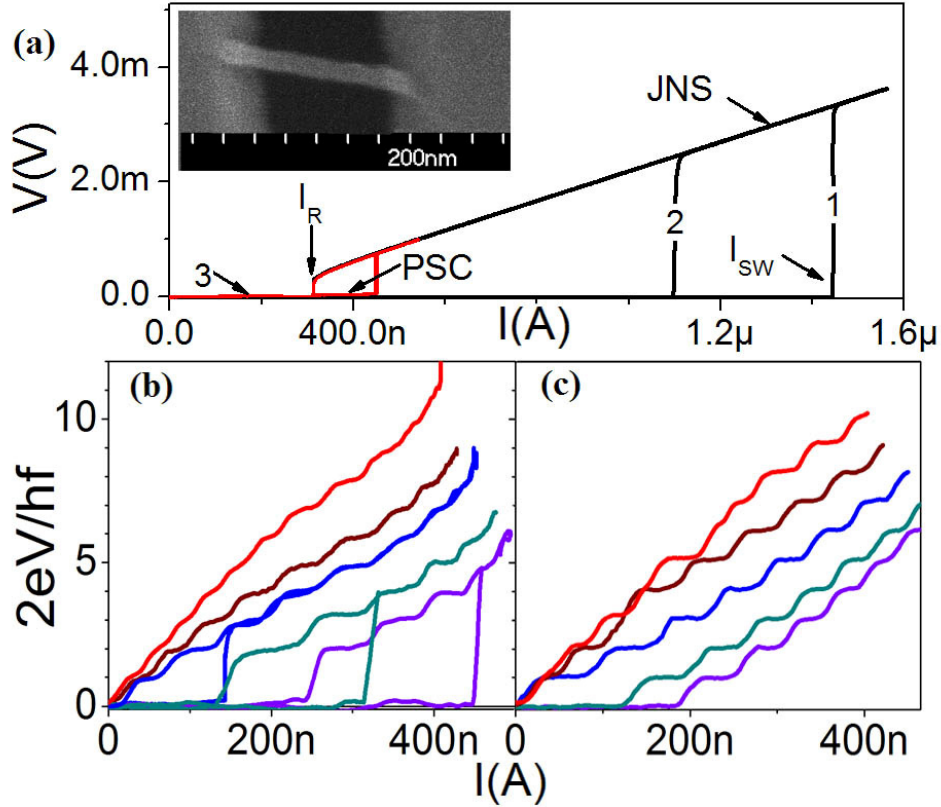

Figure 1. (a) (color online) Positive bias, *V(I)* curves for Sample A taken at T=290 mK with the switching current, $I_{SW}$, and the return current, $I_R$, indicated with arrows. The JNS and PSC regimes are also indicated by arrows. The curve 1 (black) is measured at zero MW power, the curves 2 (black) and 3 (red) are measured at -31dBm and -21dBm output MW power, at 3GHz frequency. Inset: SEM image of sample A. (b) Normalized Voltage, *2eV/hf vs. I* curves for the PSC regime in sample A for applied MW powers decreasing from top to bottom curves as follows (in dBm): -21.8, -22.4, -22.9, -24, -24.6. These curves end where the wire switches to JNS. The curves measured at lower power MW show hysteresis and do not show the n=1 step, since they remain fully superconducting up to a higher current. (c) Numerical simulations using the CPR given in the text for *L*=140nm, ξ=10nm, Ω=0.1, *T*=0.3K, $I_C$ = 500 nA, for $i_{ac}$ decreasing from top curve to bottom curve as follows: 2.4, 2.0, 1.6, 0.8, 0.6. The simulated curves are extremely sensitive to the values of ξ and Ω used but show a similar trend to the experimental ones.

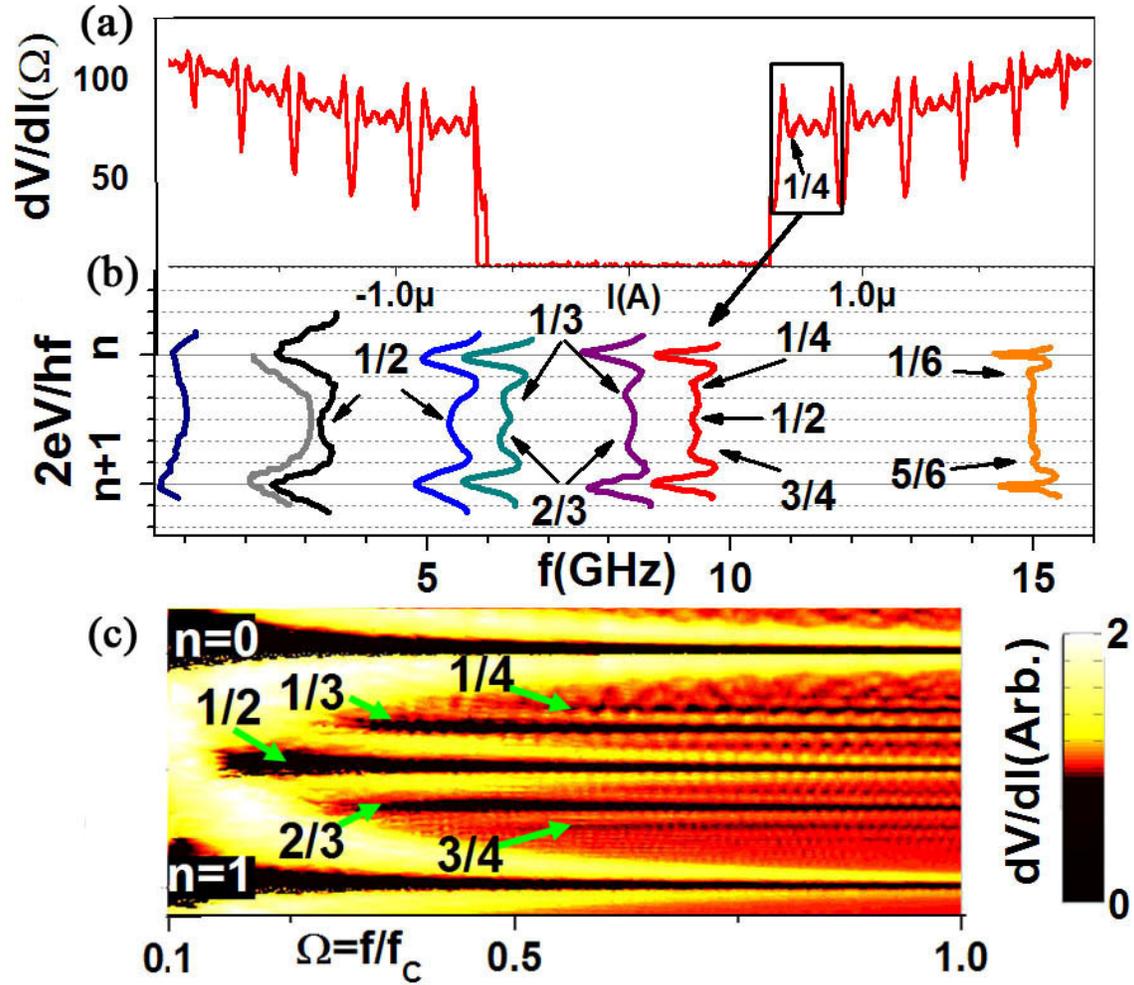

Figure 2. (a) *dV/dI*(*I*) for sample B taken at 9.5 GHz and 500 mK showing clear ¼ resonances. The boxed region represents the portion of the curve located between steps n=3 and n=4. (b) The boxed region of the curve is shown here for various microwave frequencies. The horizontal axis is *dV/dI* (in arbitrary units). Each curve is shifted to line up with the MW frequency at which each curve was measured. The vertical axis is normalized voltage, *2eV/hf*. The frequencies starting from the left are (in GHz): 0.9, 2.7, 2.9, 5.4, 6.2, 8.2, 9.5, 15. The appearance of fractional steps is indicated by arrows. (c) Numerical simulation for all the resonances between the n=0 and n=1 step using the CPR given by equation (2) as a function of the normalized frequency, $\Omega=f/f_C$ taken at $i_{ac}=1$ and $L/\xi$=130 nm/6 nm.

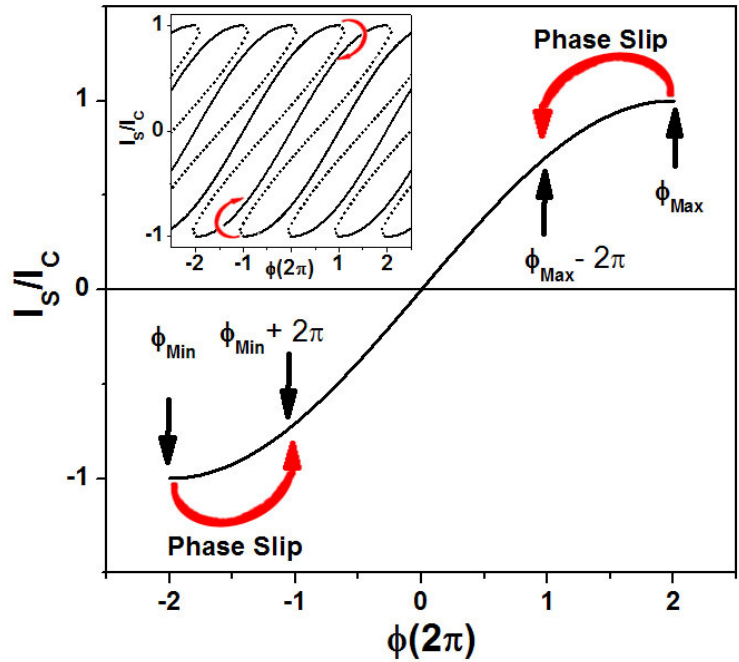

Figure 3. Schematic of the CPR used in numerical simulations. The CPR is shown as the black line, which represents supercurrent plotted versus the phase difference between the ends of the wire. When the supercurrent in the wire hits $I_C$, a phase slip occurs and the phase changes by $\pm 2\pi$ (such changes are shown by the red arrows). After the phase slip process occurs the amount of supercurrent in the wire is given by the CPR evaluated at the resulting new phase difference value, as indicated by the red arrows. Inset: Schematic, multi-valued representation of the same CPR. The solid curves are the stable branches described by the equation given in the text, each separated by a phase difference of $2\pi$. The dotted line shows the unstable branches, which are not used in our simulation. Here the red arrows show how a phase slip is equivalent to moving to the adjacent stable branch instead of traversing the unstable region.

31. Thermal fluctuations are accounted for by adding a random term $2e/\hbar \sigma \sqrt{2k_B TtR}$ to the phase evolution computation in each time step (t). Here T is the temperature and σ is a Gaussian random variable, with $\langle \sigma^2 \rangle = 1$.